\documentclass{elsart}
\usepackage{graphicx,natbib,amssymb}
\journal{Physica A}

\begin{document}
\begin{frontmatter}
\title{A Generalized Preferential Attachment Model for
 Business Firms Growth Rates:\\I. Empirical Evidence}

\author[1,2]{Fabio~Pammolli\corauthref{cor}},
\corauth[cor]{Corresponding author.} \ead{pammolli@gmail.com}
\author[3]{Dongfeng~Fu},
\ead{dffu@buphy.bu.edu}
\author[4]{S.~V.~Buldyrev},
\ead{buldyrev@yu.edu}
\author[1]{Massimo~Riccaboni},
\ead{riccaboni@unifi.it}
\author[3]{Kaushik~Matia},
\ead{kaushik@buphy.bu.edu}
\author[5]{Kazuko~Yamasaki},
\ead{yamasaki@rsch.tuis.ac.jp}
\author[3]{H.~E.~Stanley\thanksref{MERCK}}
\ead{hes@buphy.bu.edu}
\thanks[MERCK]{The Merck Foundation is gratefully acknowledged for financial support.}

\address[1]{Faculty of Economics,
  University of Florence, Via delle Pandette 9, Florence, 50127 Italy}
\address[2] {IMT Institute for
  Advanced Studies, Via S.~Micheletto 3, Lucca, 55100 Italy}
\address[3]{Center for Polymer Studies and Department of Physics, Boston
  University, Boston, MA 02215 USA }
\address[4] {Department of
 Physics,~Yeshiva University, 500 West 185th Street,~New York, NY 10033 USA}
\address[5] {Tokyo University of Information Sciences, Chiba City 265-8501
 Japan}

\begin{abstract}
  We introduce a model of proportional growth to explain the distribution
  $P(g)$ of business firm growth rates. The model predicts that $P(g)$ is
  Laplace in the central part and depicts an asymptotic power-law
  behavior in the tails with an exponent $\zeta=3$. Because of data
  limitations, previous studies in this field have been focusing exclusively
  on the Laplace shape of the body of the distribution. We
  test the model at different levels of aggregation in the economy, from
  products, to firms, to countries, and we find that the its predictions
  are in good agreement with empirical evidence on both growth distributions
  and size-variance relationships.
\end{abstract}


\end{frontmatter}

\section{Introduction}
\label{sec:Introduction}

Gibrat~\citep{Gibrat31}, building upon the work of the astronomers
Kapteyn and Uven~\citep{Kapteyn16}, assumed the expected value of
the growth rate of a business firm's size to be proportional to the
current size of the firm (the so called ``Law of Proportionate
Effect'')~\citep{Zipf49,Gabaix99}. Several models of proportional
growth have been subsequently introduced in economics to explain the
growth of business firms~\citep{Steindl65,Sutton97,Kalecki45}. Simon
and co-authors~\citep{Simon55,Simon77} extended Gibrat's model by
introducing an entry process according to which the number of firms
rise over time. In Simon's framework, the market consists of a
sequence of many independent ``opportunities'' which arise over
time, each of size unity. Models in this tradition have been
challenged by many
researchers~\citep{Stanley96,Lee98,Stanley99,Bottazzi01,Matia04,Fuetal2005}
who found that the firm growth distribution is not Gaussian but
displays a tent shape.

Using a database on the size and growth of firms and products, we
characterize the shape of the whole growth rate distribution. Then
we introduce a general framework that provides an unifying
explanation for the growth of business firms based on the number and
size distribution of their elementary constituent
components~\citep{Amaral97,Sergey_II,Sutton02,DeFabritiis03,Amaral98,Takayasu98,Canning98,Buldyrev03,Fuetal2005}.
Specifically we present a model of proportional growth in both the
number of units and their size and we draw some general implications
on the mechanisms which sustain business firm
growth~\citep{Simon77,Sutton97,Kalecki45,DeFabritiis03}. According
to the model, the probability density function (PDF) of growth rates
is Laplace in the center~\citep{Stanley96} with power law
tails~\citep{Reed01}. We test our model by analyzing different
levels of aggregation of economic systems, from the ``micro'' level
of products to the ``macro'' level of industrial sectors and
national economies. We find that the model accurately predicts the
shape of the PDF of growth rate at any level of aggregation.

\section{The Model}
\label{sec:Model}

\begin{figure}[htb]
\centering
\includegraphics[width=8cm,angle=-90]{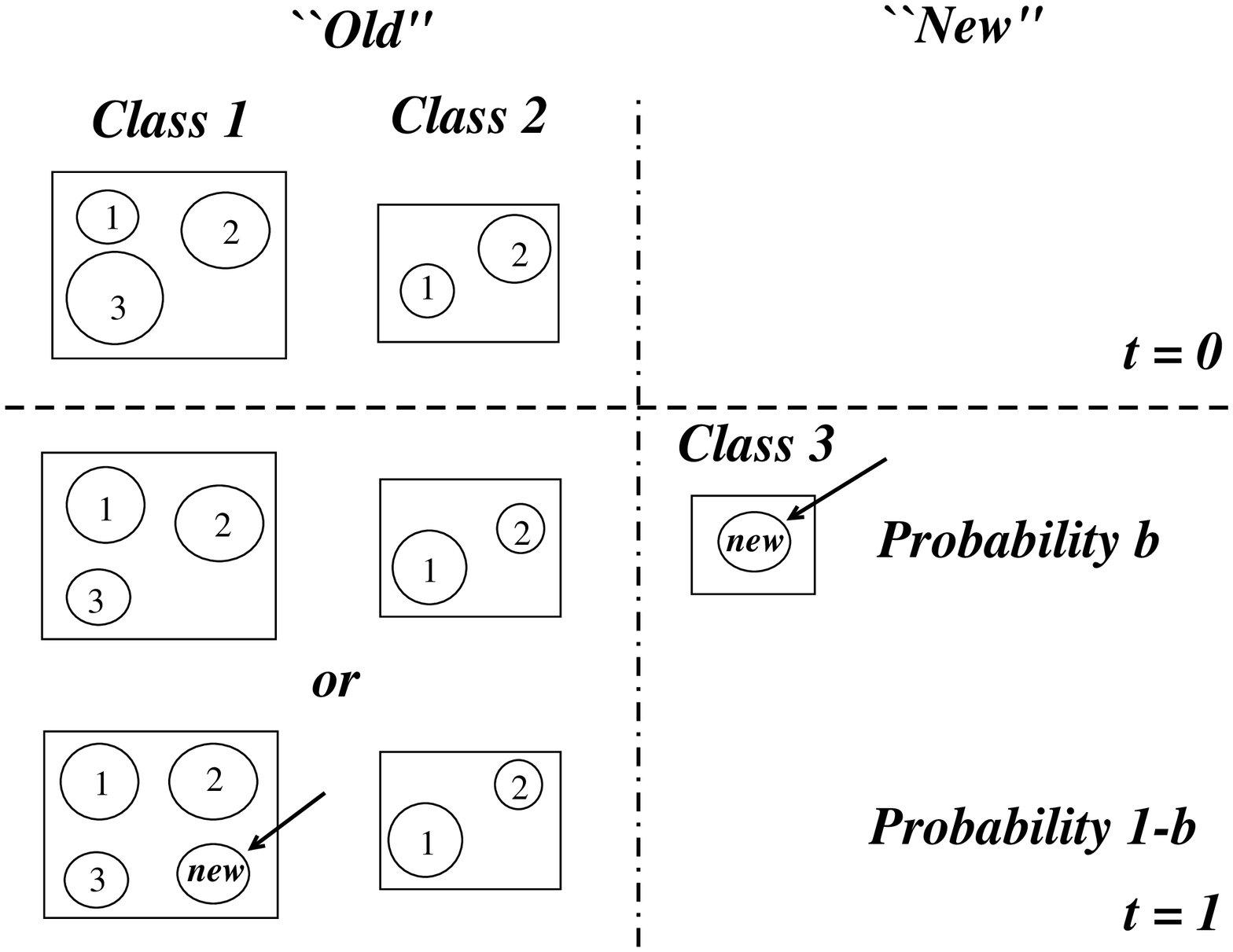}
\caption{Schematic representation of the model of proportional growth. At
time $t=0$, there are $N(0)=2$ classes ($\square$) and $n(0)=5$ units
($\bigcirc$) (Assumption A1). The area of each circle is proportional to the
size $\xi$ of the unit, and the size of each class is the sum of the areas of
its constituent units (see Assumption B1). At the next time step, $t=1$, a new unit is
created (Assumption A2). With probability $b$ the new unit is assigned to a
new class (class 3 in this example) (Assumption A3). With probability $1-b$
the new unit is assigned to an existing class with probability proportional
to the number of units in the class (Assumption A4). In this example, a new
unit is assigned to class $1$ with probability $3/5$ or to class $2$ with
probability $2/5$. Finally, at each time step, each circle $i$ grows or
shrinks by a random factor $\eta_i$ (Assumption B2).}
\label{schematic}
\end{figure}

We model business firms as classes consisting of a random number of
units. According to this view, a firm is represented as the
aggregation of its constituent units such as
divisions~\citep{Amaral98}, businesses~\citep{Sutton02}, or
products~\citep{DeFabritiis03}. We study the logarithm of the
one-year growth rate of classes $g\equiv \log(S(t+1)/S(t))$ where
$S(t)$ and $S(t+1)$ are the sizes of classes in the year $t$ and
$t+1$ measured in monetary values (GDP for countries, sales for
firms and products). The model is illustrated in
Fig.~\ref{schematic}. The model is built upon two key sets of
assumptions:
\begin{itemize}
 \item[A)] the number of units in a class grows
in proportion to the existing number of units;
 \item[B)] the size of each unit grows in proportion to its size.
\end{itemize}

More specifically, the first set of assumptions is:
\begin{enumerate}
\item[(A1)] Each class $\alpha$ consists of $K_{\alpha}(t)$ number of
  units. At time $t=0$, there are $N(0)$ classes consisting of $n(0)$ total number of units.
\item[(A2)] At each time step a new unit is created. Thus the number
  of units at time $t$ is $n(t)=n(0)+t$.
\item[(A3)] With birth probability $b$, this new unit is assigned to a new
  class.
\item[(A4)] With probability $1-b$, a new unit is assigned to an existing
  class $\alpha$ with probability $P_{\alpha}=(1-b)K_{\alpha}(t)/n(t)$.
\end{enumerate}

The second set of assumptions of the model is:

\begin{enumerate}
\item[(A5)] At time $t$, each class $\alpha$ has $K_{\alpha}(t)$ units of
size $\xi_i(t)$, ${i=1,2,...K_{\alpha}(t)}$ where $K_{\alpha}$ and
$\xi_i > 0$ are independent random variables.
\item[(A6)] At time $t+1$, the size of each unit is decreased or
  increased by a random factor $\eta_i(t)>0$ so that
\begin{equation}
\xi_i(t+1)=\xi_i(t)\,\eta_i(t),
\end{equation}
where $\eta_i(t)$, the growth rate of unit $i$, is independent
random variable.

\end{enumerate}

Based on the first set of assumptions, we derive $P(K)$, the
probability distribution of the number of units in the classes at
large $t$. Then, using the second set of assumptions with $P(K)$ we
calculate the probability distribution of growth rates $P(g)$. Since
the exact analytical solution of $P(K)$ is not known, we provide
approximate mean field solution for $P(K)$ (see, e.g., Chapter 6
of~\citep{book}). We also assume that $P(K)$ follows exponential
distribution either in old and new classes~\citep{Cox}.

Therefore, the distribution of units in all classes is given by
 \begin{equation}
 P(K)= \frac{N(0)}{N(0) + bt}\,P_{old}(K)\,+\,\frac{bt}{N(0)+bt}\,P_{new}(K).
 \label{p_K_final}
 \end{equation}
where $P_{old}(K)$ and $P_{new}(K)$ are the distribution of units in
pre-existing and new classes, respectively.

Let us assume both the size and growth of units ($\xi_i$ and $\eta_i$
respectively) are distributed as $LN(m_\xi, V_\xi)$ and $LN(m_\eta, V_\eta)$
where $LN$ means lognormal distribution. Thus, for large $K$, $g$ has a
Gaussian distribution
\begin{equation}
P(g|K)={\frac{\sqrt K}{\sqrt{2\pi V}}}\,\exp\left(-\frac{(g-m)^2K}{2V}\right),
\label{P_large_K}
\end{equation}
where $m$ is the function of $m_\eta$ and $V_\eta$, and $V$ is the function
of $V_\xi$ and $V_\eta$. Thus, the resulting distribution of the growth rates
of all classes is determined by
\begin{equation}
  P(g) \equiv \sum_{K=1}^{\infty}P(K)P(g|K).
  \label{P_g_sum}
\end{equation}
The approximate solution of $P(g)$ is obtained by using
Eq.~(\ref{P_large_K}) for $P(g|K)$ for finite $K$, mean field
solution Eq.~(\ref{p_K_final}) for $P(K)$ and replacing summation by
integration in Eq.~(\ref{P_g_sum}). After some algebra, we find that
the the shape of $P(g)$ based on either $P_{old}(K)$ or $P_{new}(K)$
is same, and $P(g)$ is given as follows
\begin{equation}
\label{p_new}
P(g)\approx \frac{2V}{\sqrt{g^2+2V}\,(|g|+\sqrt{g^2+2V})^2}.
\end{equation}
which behaves for $g\to 0$ as $1/\sqrt{2V}-|g|/V$ and for
$g\to\infty$ as $V/(2g^3)$. Thus, the distribution is well
approximated by a Laplace distribution in the body with power-law
tails.

\section{The Empirical Evidence}

We analyze different levels of aggregation of economic systems, from
the micro level of products to the macro level of industrial sectors
and national economies.

We study a unique database, the pharmaceutical industry database
(PHID), which records sales figures of the $189,303$ products
commercialized by $7,184$ pharmaceutical firms in $21$ countries
from 1994 to 2004, covering the whole size distribution for products
and firms and monitoring the flows of entry and exit at both levels.
Moreover, we investigate the growth rates of all U.S.
publicly-traded firms from 1973 to 2004 in all industries, based on
Security Exchange Commission filings (Compustat). Finally, at the
macro level, we study the growth rates of the gross domestic product
(GDP) of $195$ countries from 1960 to 2004 (World Bank).

\begin{figure}[htb]
\centering
\includegraphics[scale=0.2,angle=-90]{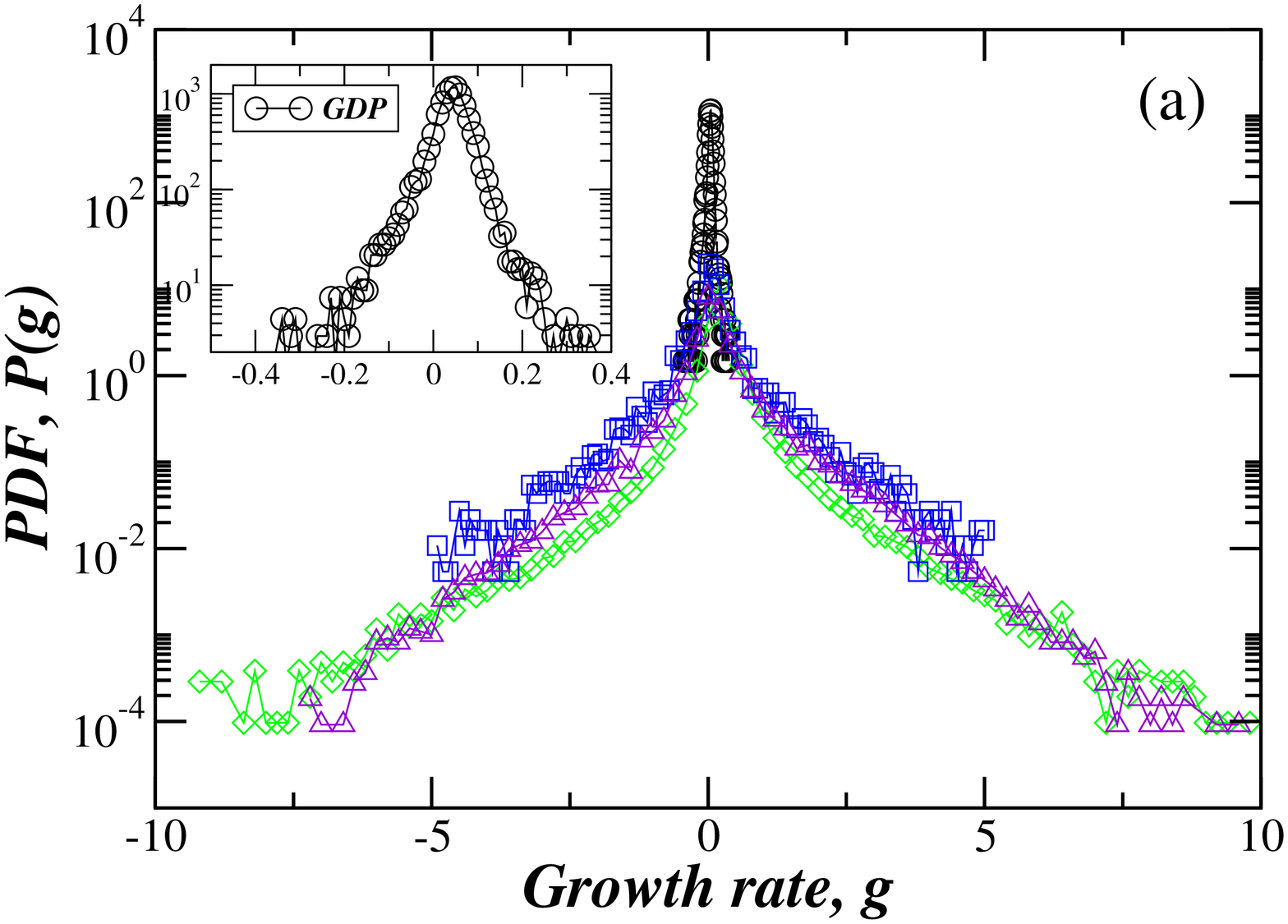}
\includegraphics[scale=0.2,angle=-90]{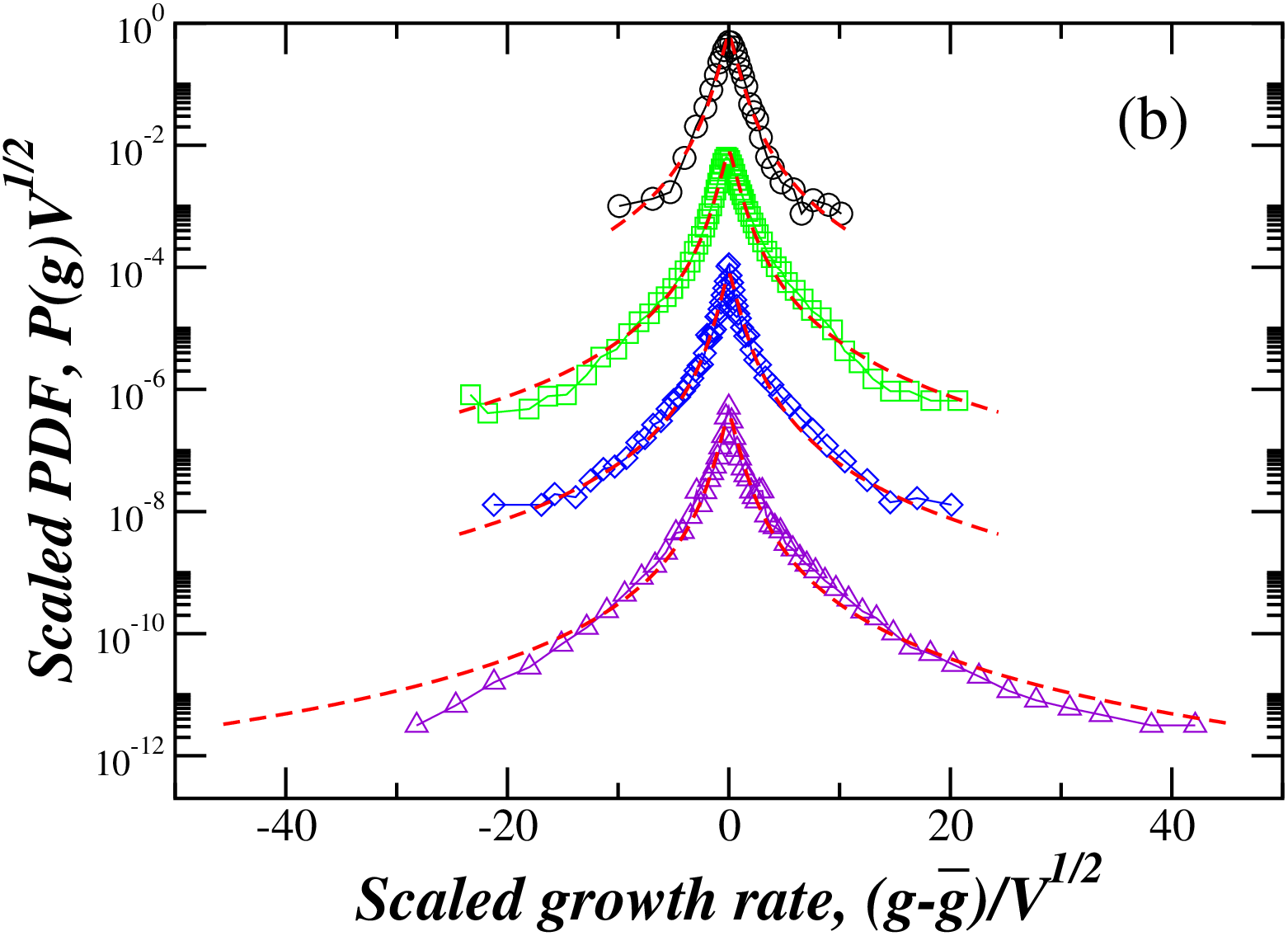}
\caption{(a) Empirical results of the probability density function (PDF)
  $P(g)$ of growth rates. Shown are country GDP ($\bigcirc$), pharmaceutical
  firms ($\square$), manufacturing firms ($\Diamond$), and pharmaceutical
  products ($\bigtriangleup$). (b) Empirical tests of
  Eq.~(\protect\ref{p_new}) for the probability density function (PDF) $P(g)$
  of growth rates rescaled by $\sqrt{V}$. Dashed lines are obtained based on
  Eq.~(\protect\ref{p_new}) with $V\approx 4\times10^{-4}$ for GDP, $V\approx
  0.014$ for pharmaceutical firms, $V\approx 0.019$ for manufacturing firms,
  and $V\approx 0.01$ for products. After rescaling, the four PDFs can be fit
  by the same function. For clarity, the pharmaceutical firms are offset by a
  factor of $10^2$, manufacturing firms by a factor of $10^4$ and the
  pharmaceutical products by a factor of $10^6$. }
\label{growth_dist_4industry}
\end{figure}

\begin{figure}[htb]
\centering
\includegraphics[scale=0.25,angle=-90]{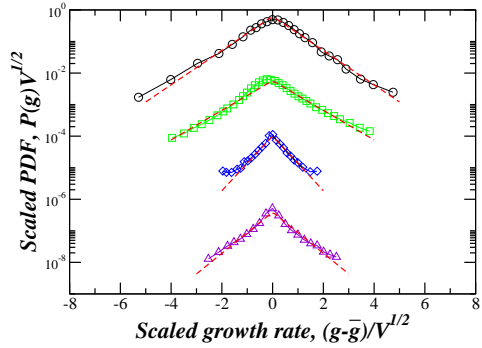}
\caption{Empirical tests of Eq.~(\protect\protect\ref{p_new}) for the {\it
central} part in the PDF $P(g)$ of growth rates rescaled by
$\sqrt{V}$. Shown are 4 symbols: country GDP ($\bigcirc$),
pharmaceutical firms ($\square$), manufacturing firms ($\Diamond$), and
pharmaceutical products ($\bigtriangleup$). The shape of central parts for
all four levels of aggregation can be well fit by a Laplace distribution
(dashed lines). Note that Laplace distribution can fit $P(g)$ only over a
restricted range, from $P(g) = 1$ to $P(g)\approx 10^{-1}$. }
\label{body}
\end{figure}

\begin{figure}[htb]
\centering
\includegraphics[scale=0.25,angle=-90]{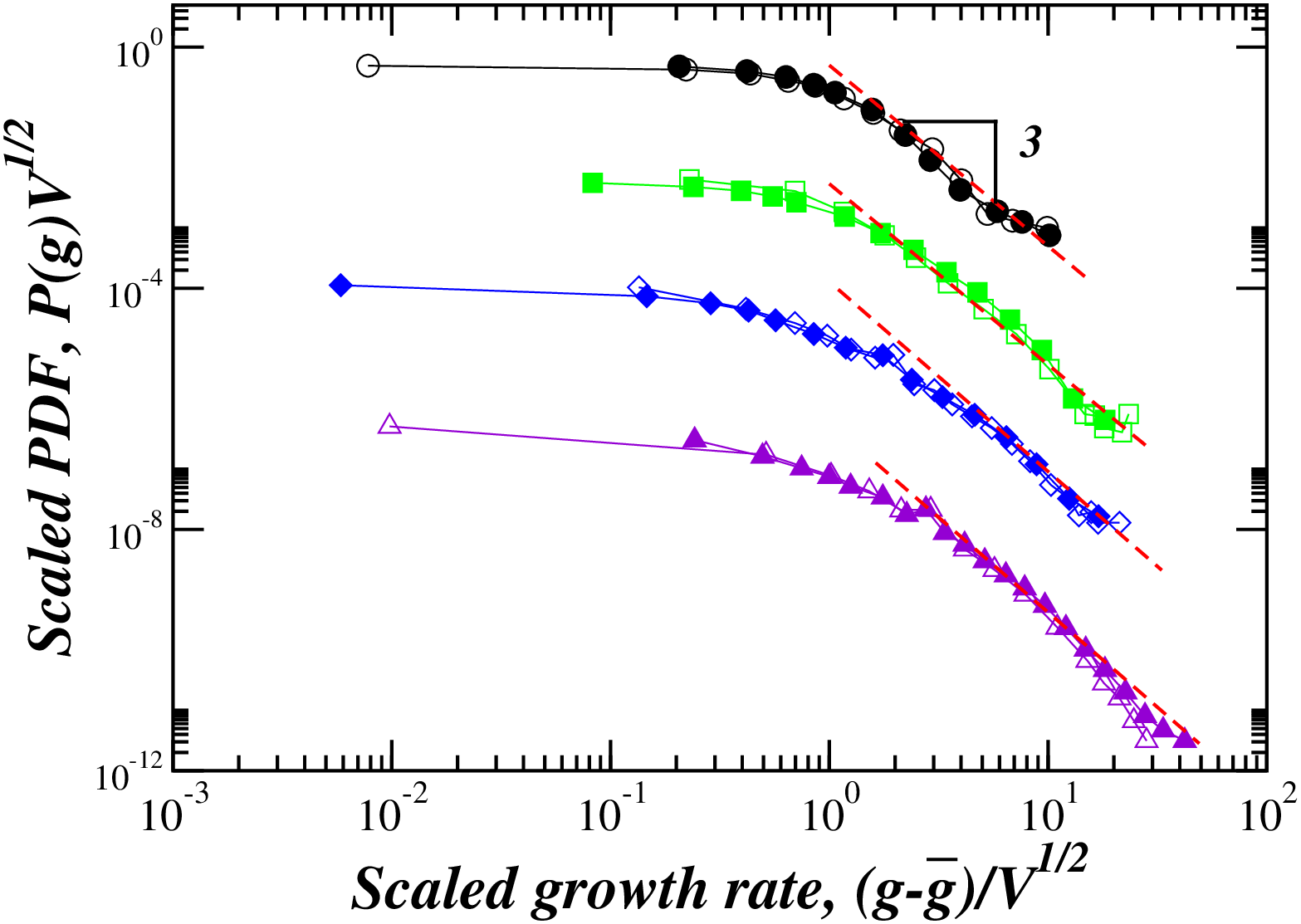}
\caption{Empirical tests of Eq.~(\protect\ref{p_new}) for the {\it tail}
parts of the PDF of growth rates rescaled by $\sqrt{V}$. The asymptotic
behavior of $g$ at any level of aggregation can be well approximated by power
laws with exponents $\protect\zeta\approx 3$ (dashed lines). The symbols are
as follows: Country GDP (left tail: $\bigcirc$, right tail: $\bullet$),
pharmaceutical firms (left tail: $\square$, right tail: $\blacksquare$),
manufacturing firms (left tail: $\Diamond$, right tail: $\blacklozenge$),
pharmaceutical products (left tail: $\bigtriangleup$, right tail:
$\blacktriangle$).} \label{tails}
\end{figure}

Figure~\ref{growth_dist_4industry}a shows that the growth
distributions of countries, firms, and products seems quite
different but in Fig.~\ref{growth_dist_4industry}b they are all well
fitted by Eq.~(\protect\ref{p_new}) just with different values of
$V$. Growth distributions at any level of aggregation depict marked
departures from a Gaussian shape. Moreover, while the $P(g)$ of GDP
can be approximated by a Laplace distribution, the $P(g)$ of firms
and products are clearly more leptokurtic than Laplace. Coherently
with the predictions of the model outlined in Section
\ref{sec:Model}, we find that both product and firm growth
distributions are Laplace in the body (Fig.~\ref{body}), with
power-law tails with an exponent $\zeta=3$ (Fig.~\ref{tails}).

%======================================================================
\section{Discussion}
We introduce a simple and general model that accounts for both the
central part and the tails of growth distributions at different
levels of aggregation in economic systems. In particular, we show
that the shape of the business firm growth distribution can be
accounted for by a simple model of proportional growth in both
number and size of their constituent units. The tails of growth rate
distributions are populated by younger and smaller firms composed of
one or few products while the center of the distribution is shaped
by big multi-product firms. Our model predicts that the growth
distribution is Laplace in the central part and depicts an
asymptotic power-law  behavior in the tails. We find that the
model's predictions are accurate.

%%%%%%%%%%%%%%%%%%%%%%%%%%%%%%%%%REFERENCE%%%%%%%%%%%%%%%%%%%%%%%%%%%%%%%%%%%%%%%%

%%%%%%%%%%%%%%%%%%%%%%%%%%%%%%%%%%%END %%%%%%%%%%%%%%%%%%%%%%%%%%%%%%%%%%%%%

\begin{thebibliography}{99}

%\bibitem{Gibrat30} Gibrat,~R. (1930) {\it Bulletin de Statistique
%G\'en\'eral, France}, {\bf 19}, 469.

\bibitem{Gibrat31} Gibrat,~R. (1931) \textit{Les In\'egalit\'es
\'{E}conomiques} (Librairie du Recueil Sirey, Paris).

\bibitem{Kapteyn16} Kapteyn,~J. \& Uven~M.~J. (1916) \textit{Skew Frequency
Curves in Biology and Statistics} (Hoitsema Brothers, Groningen).

\bibitem{Zipf49} Zipf,~G. (1949) {\it Human Behavior and the Principle of
Least Effort} (Addison-Wesley, Cambridge, MA).

\bibitem{Gabaix99} Gabaix,~X. (1999) {\it Quar. J. Econ.} {\bf 114},
739--767.

\bibitem{Steindl65} Steindl,~J. (1965) \textit{Random Processes and the Growth of
Firms: A study of the Pareto law} (London, Griffin).

%%%%%%%%%%%%%%%%%%%%%%%%%% growth rate P(g) and P(S) %%%%%%%%%%%%%%%%%%
\bibitem{Sutton97} Sutton,~J. (1997) {\it J. Econ. Lit.} {\bf 35}, 40-59.

\bibitem{Kalecki45} Kalecki,~M. (1945) \textit{Econometrica} \textbf{13},
161-170. %On the Gibrat Distribution

\bibitem{Simon55} Simon,~H.~A. (1955) {\it Biometrika}, {\bf 42}, 425-440.

%\bibitem{Simon58} Simon,~H.~A. \& Bonini,~C.~P. (1958) {\it Am. Econ. Rev.}
%{\bf 48}, 607-617.
%
%\bibitem{Simon75} Ijiri,~Y. \& Simon,~H.~A. (1975) {\it
%Proc. Nat. Acad. Sci.} {\bf 72}, 1654-1657.

\bibitem{Simon77} Ijiri,~Y. \& Simon,~H.~A., (1977) {\it Skew distributions
and the sizes of business firms} (North-Holland Pub. Co., Amsterdam).

%%%%%%%%%%%%%%%%%%%%%%% Stanley etc's findings %%%%%%%%%%%%%%%%%%%
\bibitem{Stanley96} Stanley,~M.~H.~R., Amaral,~L.~A.~N., Buldyrev,~S.~V.,
Havlin,~S., Leschhorn,~H., Maass,~P., Salinger,~M.~A. \&
Stanley,~H.~E. (1996) {\it Nature} {\bf 379}, 804-806.

\bibitem{Lee98} Lee,~Y., Amaral,~L.~A.~N., Canning,~D., Meyer,~M. \&
Stanley,~H.~E.~ (1998) {\it Phys. Rev. Lett.} {\bf 81}, 3275-3278.

\bibitem{Stanley99} Plerou,~V., Amaral,~L.~A.~N., Gopikrishnan,~P.,
Meyer,~M. \& Stanley,~H.~E. (1999) {\it Nature} {\bf 433}, 433-437.

\bibitem{Bottazzi01} Bottazzi,~G., Dosi,~G., Lippi,~M., Pammolli,~F. \&
Riccaboni,~M. (2001) \textit{Int. J. Ind. Org.} \textbf{19}, 1161-1187.

\bibitem{Matia04} Matia,~K., Fu,~D., Buldyrev,~S.~V., Pammolli,~F.,
Riccaboni,~M. \& Stanley,~H.~E. (2004) {\it Europhys. Lett.} {\bf
67}, 498-503.

\bibitem{Fuetal2005} Fu,~D., Pammolli,~F., Buldyrev,~S.V.,
Riccaboni,~M., Matia,~K., Yamasaki,~K., Stanley,~H.E. (2005) {\it
PNAS} {\bf 102}, 18801--18806.


%%%%%%%%%%%%%%%%%%%%%%%%% Stanley's Model %%%%%%%%%%%%%%%%%%%%%%%%%%%%%%%%%
\bibitem{Amaral97} Amaral,~L.~A.~N., Buldyrev,~S.~V., Havlin,~S.,
Leschhorn,~H, Maass,~P., Salinger,~M.~A., Stanley,~H.~E. \&
Stanley,~M.~H.~R. (1997) {\it J. Phys. I France} {\bf 7}, 621--633.

\bibitem{Sergey_II} Buldyrev,~S.~V., Amaral,~L.~A.~N., Havlin,~S.,
Leschhorn,~H, Maass,~P., Salinger,~M.~A.~, Stanley,~H.~E. \&
Stanley,~M.~H.~R. (1997) {\it J.~Phys.~I France} {\bf 7}, 635--650.

\bibitem{Sutton02} Sutton,~J. (2002) {\it Physica A} {\bf 312}, 577--590.

\bibitem{DeFabritiis03} Fabritiis,~G.~D., Pammolli,~F. \&
Riccaboni,~M. (2003) {\it Physica A} {\bf 324}, 38--44.

\bibitem{Amaral98} Amaral,~L.~A.~N., Buldyrev,~S.~V., Havlin,~S.,
Salinger,~M.~A. \& Stanley,~H.~E. (1998) {\it Phys. Rev. Lett} {\bf
80}, 1385--1388.

\bibitem{Takayasu98} Takayasu,~H. \& Okuyama,~K. (1998) {\it Fractals\/} {\bf
6}, 67--79.

\bibitem{Canning98} Canning,~D., Amaral,~L.~A.~N., Lee,~Y., Meyer,~M. \&
Stanley,~H.~E. (1998) {\it Econ.~Lett.} {\bf 60}, 335--341.

\bibitem{Buldyrev03} Buldyrev,~S.~V., Dokholyan,~N.~V., Erramilli,~S.,
  Hong,~M., Kim,~J.~Y., Malescio,~G. \& Stanley,~H.~E. (2003) {\it Physica A}
  {\bf 330}, 653--659.

%%%%%%%%%%%%%%%%%%%%%%%%%%%%% p(S) and P(g) %%%%%%%%%%%%%%%%%%%%%%%%%%%%%%%%%%

\bibitem{Kalecki}
Kalecki,~M.~R. {\it Econometrica\/} (1945) {\bf 13}, 161--170.

%"Firm Size and Rate of Growth."
%\bibitem{Mansfield} Mansfield,~D.~E. (1962) {\it Am.~Econ.~Rev.} {\bf 52},
%1024--1051.
%
%\bibitem{Hall}Hall,~B.~H. (1987) {\it J.~Ind.~Econ.} {\bf 35}, 583--606.

%%%%%%%%%%%%%%%%%%%%%% power law tails %%%%%%%%%%%%%%%%%%%%%%%%%%%%%%%%%%%%%%
\bibitem{Reed01} Reed,~W.~J. (2001) {\it Econ. Lett.} {\bf 74}, 15--19.

%\bibitem{Reed02} Reed,~W.~J. \& Hughes,~B.~D. (2002) {\it Phys. Rev. E} {\bf
%66}, 067103.

%"Markov-Perfect Industry Dynamics: a Framework for Empirical Work."

%%%%%%%%%%%%%%%%%%Laplace Dist. %%%%%%%%%%%%%%%%%%%%%%%%%%%%%%%%%%%%%%%%%%%%%%%
%%%%%%%%%%%%%%%%%%% solution of urn model %%%%%%%%%%%%%%%%%%%%%%%%%%%%%%%%

\bibitem{book} Stanley,~H.~E. (1971) {\it Introduction to Phase Transitions and
Critical Phenomena} (Oxford University Press, Oxford).

\bibitem{Cox} Cox,~D.~R. \& Miller,~H.~D. (1968) {\it The Theory of
Stochastic Processes} (Chapman and Hall, London).

\bibitem{Kotz01} Kotz,~S., Kozubowski,~T.~J. \& Podg\'orski,~K. (2001) {\it
The Laplace Distribution and Generalizations: A Revisit with Applications to
Communications, Economics, Engineering, and Finance} (Birkhauser, Boston).

%%%%%%%%%%%%%%%%%%%%%%%%%%%%%%%%%%%%%%%%%%%%%%%%%%%%%%%%%%%%%%%%%%%%%%%%%%

\end{thebibliography}
\end{document}